\documentclass[12pt]{iopart}

\usepackage[draft]{hyperref}
\usepackage{iopams}
\usepackage{epsfig}
\usepackage{cite}

\newcommand{\be}{\begin{equation}}
\newcommand{\ee}{\end{equation}}
\newcommand{\ba}{\begin{eqnarray}}
\newcommand{\ea}{\end{eqnarray}}
\newcommand{\bra}[1]{\ensuremath{\langle #1 |}}
\newcommand{\ket}[1]{\ensuremath{|#1\rangle}}

\newcommand{\om}{\ensuremath{\omega}}

\newcommand{\ave}[1]{\ensuremath{\langle #1\rangle}}
\renewcommand{\a}{\ensuremath{\hat a}}

\newcommand{\Em}{\ensuremath{\hat E^{(-)}}}
\newcommand{\Ep}{\ensuremath{\hat E^{(+)}}}
\newcommand{\ad}{\ensuremath{\hat a^{\dagger}}}
\newcommand{\bd}{\ensuremath{\hat b^{\dagger}}}

\begin{document}

\title[Accessing the purity of a single photon by the width of the HOM interference]{Accessing the purity of a single photon by \\ the width of the Hong-Ou-Mandel interference}

\author{Kati\'uscia N. Cassemiro$^1$, Kaisa Laiho$^1$ and Christine Silberhorn$^{1,2}$}
\address{1 Max Planck Institute for the Science of Light, G\"{u}nther-Scharowsky-Str. 1 / Bau 24, 91058 Erlangen, Germany.}
\address{2 University of Paderborn, Applied Physics, Warburgerstrasse 100, 33098 Paderborn, Germany.}

\ead{Katiuscia.Cassemiro@mpl.mpg.de}


\begin{abstract}
We demonstrate a method to determine the spectral purity of single photons. The technique is based on the Hong-Ou-Mandel (HOM) interference between a single photon state and a suitably prepared coherent field. We show that the temporal width of the HOM dip is not only related to reciprocal of the spectral width but also to the underlying quantum coherence. Therefore, by measuring the width of both the HOM dip and the spectrum one can directly quantify the degree of spectral purity. The distinct advantage of our proposal is that it obviates the need for perfect mode matching, since it does not rely on the visibility of the interference. Our method is particularly useful for characterizing the purity of heralded single photon states. 
\end{abstract}

\maketitle

\section{Introduction}
The Hong-Ou-Mandel interference is one of the most famous phenomenon studied in quantum optical experiments~\cite{hom}.  The effect can be observed by coherently combining two indistinguishable photons at a beam splitter. As a result of interference in the Hilbert space the photons bunch together and depart at the same beam splitter output port. Therefore, the effect relies not only on the linear superposition of electric fields, but it also highlights the specific quantum characteristic of the input states.

Originally, the HOM effect was observed by interfering two strongly correlated fields produced by parametric down-conversion~\cite{hom,ou_prl67,kwiat_pra45}. 
High visibility was achieved by tight spectral and spatial filtering due to the incoherent nature of the light fields. Nowadays the two-photon coalescence effect constitutes the core mechanism of many quantum information protocols, such as linear-optics quantum computation~\cite{flamme} and teleportation~\cite{bennett_prl70,bouwmeister}. This fact has boosted the efforts towards the experimental generation of pure single photon states~\cite{louinis}, culminating in the observation of HOM interference between single photons emitted by a semiconductor quantum dot~\cite{santoro} or dissimilar sources~\cite{shields_nphys}. 

Given the current demand on preparing high quality single photon states as resource for quantum information tasks, the experimental determination of their degree of purity and underlying spectral properties becomes paramount. At present, the standard adopted benchmark is the visibility of the HOM interference. However, since it indicates the degree of indistinguishability between the quantum states~\cite{mosley_njp, sun_pra}, it renders a pessimistic estimate for the photon purity as it is highly sensitive to mode mismatching. Although a complete mapping of the spectral properties could be realized~\cite{wasilewsky, wasilewsky2}, we are here concerned with the characterization of a single quantifier, i.e. the degree of purity.

As already noticed in the seminal Hong-Ou-Mandel paper~\cite{hom}, the fourth-order interference experiment provides a way to determine properties of the interfering photons with extremely high temporal resolution, on the order of femtoseconds. In the earlier experiments, the temporal width of the dip was mainly determined by the inverse bandwidth of the employed spectral filters, which was indeed equal to the coherence time.  This fact was just a consequence of the choice of input state--an incoherent parametric down-converted one. In a more general situation the width  of the HOM dip can greatly differ from the coherence time.
Still, the properties of  the temporal width of the HOM interference have been mostly ignored. We show here that the latter is related to both the reciprocal of the spectral width and  the underlying quantum coherence, thus providing crucial knowledge about the state's spectral purity.
 
In this paper we present a method to characterize the degree of spectral purity in a robust way and with few measurements. In order to do that we analyze the temporal profile of the HOM interference between a single photon state and an auxiliary weak coherent field, used as a reference. The advantage of our proposal is that it enables the quantification of the purity without the need for perfect mode matching, thus greatly reducing the technical demands on the experimental conditions. In the simplest case, our method requires only the knowledge of the spectral widths of the reference and single photon state, together with the temporal width of the HOM dip. The first two quantities are in general easily measured by standard high-sensitive spectrographs~\cite{migdall, malte_optlet}.

Our technique gives exact results when two assumptions commonly found in the literature are made. Moreover, it provides good results even in the most general case.
Our proposal is well adapted for photon pair sources, which we discuss theoretically, showing the expected performance under different conditions. The best situation occurs when employing a practical and popular strategy to generate pure states, the one relying on tight spectral filtering.  For other situations an error limited to about $10\%$ may occur in a worse case scenario. To provide concrete results that illustrate the applicability of our technique, we also present an experimental investigation. Our findings corroborate the usefulness and benefits of the method.

This paper is organized as follows. In section~\ref{secpurity} we present general results relating the width of the HOM dip with the spectral purity of the single photon state. We define the photon {\it spectral density function} and show that in a Gaussian scenario it can be parametrized by two widths. While the major one is the spectral width,  the minor one is related to the reciprocal of the HOM time duration. The purity acquires a simple geometrical interpretation, and is shown to be the ratio between those two widths. The first order coherence time, on the other hand,  is the inverse of their difference. Therefore, as well known, spectrally pure states have a coherence time that tends to infinity. Impurity, by contrast, leads to a broader spectrum, thus shorter coherence time, resulting in a wavepacket that is not Fourier-transform-limited. After establishing those concepts we introduce our method for characterizing the purity. In section~\ref{secapplication} we discuss the validity of our proposal when applied to states arising from photon pair sources. Section~\ref{secexperiment} covers an experimental investigation, where we employ our technique to characterize the spectral purity of  signal and idler photons generated in a wave\-guided parametric down-conversion source. Further theoretical details about photon pair sources is provided in the Appendix.

\section{Purity, Temporal Coherence and HOM Interference}
\label{secpurity}
\begin{figure}[ht]
\centering
\epsfig{file=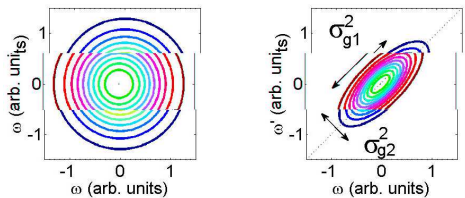,scale=0.35}
\caption{Contour lines of the density function $|g(\om , \om ')|$. Left: Pure state with Gaussian spectral profile. Right: Pictorial representation of an impure  spectral density function, with major and minor widths denoted by $\sigma_{g1}$ and $\sigma_{g2}$, respectively. This particular form of $g$ is separable along the $\pm 45^{\circ}$ axis.}
\label{gww}
\end{figure}

Before explaining how to retrieve the spectral purity from the HOM interference, we first define a pure single photon state prepared with a spectral shape $u(\om)$ as~\cite{raymer}
\be
\ket{\Psi}=\int d\om \, u^*(\om) \, \bd(\om) \ket{0}\,.
\ee
From its corresponding density operator one observes that the spectral density function is simply $g_{\mathrm{pure}}(\om,\om')=u^*(\om)u(\om')$, which is symmetric with respect to $\om$ and $\om'$. Considering a field with a Gaussian spectral shape the function $g_{\mathrm{pure}}$ presents a circular symmetry in frequency space,  as illustrated in figure~\ref{gww} (left).
To include the description of impure states, a more general representation for the single photon state is used
\be
\rho_s =  \int \int d\om \; d\om' \; g(\om,\om') \; \ad(\om)\ket{0}\bra{0}\a(\om')\,.
\ee
Since density operators are Hermitian, any spectral density function $g$ is symmetric about the 45$^{\circ}$ axis, i.e. about $(\om +\om')$.  As usual, the diagonal elements of the density matrix correspond to the {\it populations}. In the present case those compose the {\it spectrum} of the field, which is directly measured with a spectrograph and whose width we denote by $\sigma_{g1}$, as depicted in figure~\ref{gww} (right). As a final remark, we recall that the degree of purity is calculated as
\be
\mathcal{P}_s= \mathrm{Tr}(\rho_s^2)=\int \int d\om \; d\om' \; |g(\om,\om')|^2 \, .
\label{purity}
\ee
Therefore, the shape of $|g(\om,\om')|$ offers direct intuitive insight into the purity of the quantum state. In particular, if $|g(\om,\om')|$ is a 2-dimensional Gaussian function the purity is just the ratio between minor an major axis, as derived later.

To provide further intuition on those physical concepts, we can also calculate the first order correlation function~\cite{glauber}
\be
G^{(1)}(\om,\om')=\mathrm{Tr}[\rho_s \ad_s (\om) \a (\om')]=g(\om,\om')\,.
\ee
We realize that the spectral density function is equal to the non-normalized first order frequency correlation function. The condition of purity as the separability of $g$ in the $\{\om,\om' \}$ frequency space corresponds to the condition of perfect first order coherence as defined by Glauber~\cite{glauber}. Thus, the techniques for tailoring the spectral properties of single photons also apply for engineering wavepackets with a specific temporal coherence~\cite{torres_optlett}.

\begin{figure}[ht]
\centering
\epsfig{file=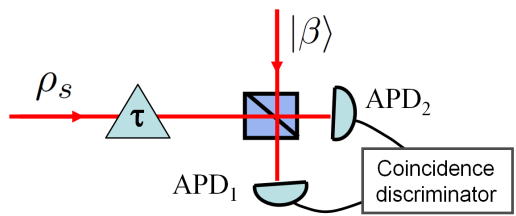,scale=0.28}
\caption{Scheme for realizing the Hong-Ou-Mandel interference between two fields with a relative optical delay $\tau$. APD: Avalanche photodetector }
\label{setuphom}
\end{figure}

The method we propose for characterizing the spectral purity $\mathcal{P}_s$ relies on the Hong-Ou-Mandel interference between a single photon state $\rho_s$ (called signal) and a highly attenuated coherent state $\ket{\beta}$ with a spectral shape $u(\om)$ (called reference)~\cite{rarity_hom, pittman_hom, konrad_hom}. The experiment is schematically shown in figure~\ref{setuphom}. Both fields interfere in a 50/50 beam-splitter, after which avalanche photo diodes (APDs) record the number of click events. The measurement of coincident events versus the time delay $\tau$ between signal and reference fields reveals the HOM interference dip. The reference field can be provided by any independent laser, since  there is no need for phase stabilization in this kind of interference.

Given that the coherent state $\ket{\beta}$ contains higher photon number contributions, to calculate the probability of coincidences we consider the weak power regime in which the interference between the reference and signal comes mostly from single photons in each mode. So far we have ignored the photon number purity of the signal state. Generally, this one is well characterized by the second order co\-he\-rence~\cite{kimble,paul_rmp}, which can be measured with a Hanbury-Brown-Twiss interferometer. Particularly for photon pair sources, it is known that the prepared state presents a small probability of two photons. We include the degrading effect of multiphoton components in a first order approximation, i.e. they are assumed not to interfere resulting only in background events.

The joint probability of registering photons at detectors $\mathrm{APD}_1$ and $\mathrm{APD}_2$ at times $t_1$ and $t_2$ is given by
\be
P_{12}(t_1,t_2)=\ave{\Em_1(t_1)\Em_2(t_2)\Ep_2(t_2)\Ep_1(t_1)}\,,
\ee
where the average is calculated over the state of the single photon source and the reference field.
The fields $\Ep_1$ and $\Ep_2$ are related to the reference $\Ep_\beta$ and signal $\Ep_s$ by the beam splitter transformation
\ba
\Ep_1(t_1)=\frac{1}{\sqrt{2}}\int_0^{\infty} \hspace{-7pt} d\omega_1 [\a_s(\omega_1)e^{-i\omega_1\tau} \hspace{-3pt} +\a_\beta(\omega_1)]e^{-i\omega_1 t_1}, \\
\Ep_2(t_2)=\frac{1}{\sqrt{2}}\int_0^{\infty} \hspace{-7pt} d\omega_2 [\a_s(\omega_2)e^{-i\omega_2\tau} \hspace{-3pt} -\a_\beta(\omega_2)]e^{-i\omega_2 t_2}.
\ea
Since the detection response time is much slower than the correlation time between the fields, we integrate $P_{12}(t_1,t_2)$ over all possible times. This leads to a probability of coincidences  given by
\be
P_c(\tau)=\mathcal{B}+\mathcal{S} [1-\mathcal{I}(\tau)] \,,
\ee
with
\be
\mathcal{I}(\tau) \hspace{-3pt}=\hspace{-3pt} \int \hspace{-2pt} \int \hspace{-2pt}d \om \, d \om '  \ u^{*}(\om ) \ g(\om , \om')  \ u(\om ')e^{i \tau( \om -\om ')} \,.
\label{interf}
\ee
The parameter $\mathcal{B}$ is a constant proportional to the probability of vacuum and two photons in both signal and reference fields, whereas $\mathcal{S}$ is proportional to the single photon contributions~\cite{kaisa_optexp}. The function $\mathcal{I}(\tau)$ accounts for the interference between the single photon components of both fields. Its value at zero delay $T=\mathcal{I}(0)$ gives the overlap between the {\it spectral density functions} of signal and reference.

The visibility of the HOM interference is defined as the fractional reduction of the probability of coincidence from its uncorrelated value, i.e.
\be
\mathcal{V}=1-\frac{P_c(\tau=0)}{P_c(\tau \rightarrow \infty)}=\frac{\mathcal{S} \,T}{\mathcal{B}+\mathcal{S}}\, .
\label{visibility}
\ee
We note in particular that, by removing background events from $P_c(\tau)$ one obtains a ``corrected'' visibility that is equal to the spectral overlap factor $T$.
Experimentally, mode mismatch in other degrees of freedom decrease even further the value of $\mathcal{V}$ from the best one $T$. We emphasize that the latter is only constrained by the spectral overlap between the density functions, which includes the property of mixedness in the signal state.

One observes from (\ref{interf}) that perfect interference is only achieved when the spectral density function $g(\om , \om ')$ is separable, i.e.  the signal photon is pure, and matches the spectrum of the reference $u(\om)$. In this special case, $\mathcal{I}(\tau)$ is nothing but the convolution of the temporal distributions of reference and signal. In a general situation, however, the width of the HOM dip will depend not only on the diagonal elements of $g(\om , \om ')$ but on the complete spectral density function. It is precisely the shape dependence of the latter with the purity that we explore.

\subsection{Method for Characterizing the Spectral Purity}
\label{secmethod}
Our aim is to find a simple relation between (\ref{purity}) and (\ref{interf}). Given the  symmetry of this problem, we analyze it in a 45$^{\circ}$ rotated frequency basis and denote by $\{x,y\}$ the new reference frame. The latter follows the direction of the major and minor axis of $g(\om,\om')$, as illustrated in figure~\ref{gww} (right). Our first assumption consists in considering that the spectral density function of the analyzed state is separable in the $\{x,y\}$ basis, thus described by $g(\om,\om')=g_1(x)\,g_2(y)$. In this case the purity and the HOM interference dip are respectively given by
\ba
\mathcal{P}_s= \int \int dx \, dy \; |g_1(x)|^2 |g_2(y)|^2  \, , \\
\mathcal{I}(\tau)= \int dx \, u^*(x) g_1(x)  \int  dy \, u(y) g_2(y) e^{i \tau y} \,.
\label{hom_simple}
\ea
We observe that the temporal shape of $\mathcal{I}(\tau)$ is only affected by the elements of the spectral density function lying at -45$^{\circ}$.  Performing the Fourier transform of the HOM dip allows one to recover the function $g_2(y)$. In particular, if  both $\mathcal{I}(\tau)$ and the spectrum of the reference are Gaussian functions, then $g_2(y)$ must also be. We recall that the spectrum of the studied signal is determined by the elements of the $g$ function lying at +45$^{\circ}$, in other words by $g_1(x)$. Therefore, the shape of the latter can be verified with an spectrograph.

Our second assumption consists in approximating the spectral density function by a Gaussian distribution and considering only linear phase terms.  This approximation is valid in many experimental situations, as we discuss later on.  Other possibility, more realistic for quantum dot sources would be to assume a Lorentzian profile. Such modification would lead to similar conclusions.  Under this condition, 
\be
g(x,y)=\frac{1}{\sqrt{2\pi \,\sigma_{g1}^2}} \, \exp\left({-\frac{x^2}{4 \, \sigma_{g1}^2} - \frac{y^2}{4 \, \sigma_{g2}^2} + i \kappa y}\right) \, ,
\ee
with $\kappa$ a constant that depends on the physical system under consideration. For a photon pair source the phase term depends only on the difference of frequencies ``$y$''. The normalization factor takes into account the condition of positivity that any physical density operator has to satisfy. With this definition of $g$, the standard deviation of the spectrum (field amplitude) is precisely $\sigma_{g1}$, and the spectral purity is given by
\be
\mathcal{P}_s=\sigma_{g2}/\sigma_{g1}\,.
\ee
Lack of purity acquires a simple geometrical interpretation: the $g$ function becomes compressed, displaying an elliptical shape in the frequency space. To characterize the purity one needs to access the values of $\sigma_{g1}$ and $\sigma_{g2}$. While we have already discussed that the first can be measured with a spectrograph, for the second one we analyze the HOM interference.

To solve (\ref{hom_simple}), we consider a reference field with a Gaussian spectrum characterized by a standard deviation $\sigma_{\beta}$. The HOM dip is determined by $\mathcal{I}(\tau)= T \; \exp{[-\tau^2/(2\delta^2)]}$, with the following overlap factor
\be
 T=\frac{1}{\delta} \; \sqrt{\frac{2}{\sigma_{g1}^2 + \sigma_{\beta}^2}} \,,
 \label{specoverlap}
\ee
and temporal width
\be
\delta^2= \frac{1}{2\,\sigma_{g2}^2}+ \frac{1}{2\,\sigma_{\beta}^2}\,.
\label{widthhomsp}
\ee
First, we emphasize that the overlap factor characterized via the width of the HOM dip quantifies exclusively the amount of {\it spectral} overlap between $u^*(\om)u(\om')$ and $g(\om,\om')$. This would not be the case when employing the visibility via equation (\ref{visibility}), given its sensitivity to mode mismatch in all degrees of freedom. Second, the width of the dip enables the characterization of $\sigma_{g2}$. Since $\sigma_{g2}=\sigma_{g1}\mathcal{P}_s$ the width of the HOM dip is not only related to reciprocal of the spectral width, as generally believed, but also to the purity of the single photon quantum state. Hence, the knowledge of very few--directly measurable--parameters completely determines the purity of the interfering photon,
\be
\mathcal{P}_s= \frac{1}{\sqrt{2 \sigma_{g1}^2 \,[\delta^2-1/(2 \sigma_{\beta}^2)]}}\,.
\label{puritywidth}
\ee

For completeness, we also calculate the coherence time $t_c$ of the single photon state. Using the generalized Wiener-Khintchine theorem~\cite{mandel}, the temporal correlation function is determined by $\gamma(t,t')=\Gamma(t,t')/\sqrt{\Gamma(t,t)\Gamma(t',t')}$, with
\be
\Gamma(t,t')=\int \int d\om \, d\om' g(\om,\om') e^{i(\om t - \om' t')}\,.
\ee
The coherence time is proportional to the standard deviation of $\gamma(t,t')$, which results in $t_c \propto 1/(\sigma_{g1}^2-\sigma_{g2}^2)$. Thus a pure state has a coherence time that tends to infinity, while incoherent chaotic sources have a coherence time that is equal to the inverse of their spectral bandwidth.

\section{Application to Photon Pair Sources}
\label{secapplication}
By employing nonlinear processes such as  three- and four-wave mixing one can generate photon pairs from an intense pump beam. The fundamental physical constraints governing these processes are energy conservation and phase-matching of pump and down-converted (signal and idler) waves. The first results in a strict photon-number correlation between the twin photons. Therefore,  one of them can be used to herald the existence of the other. Moreover, those conservation laws lead to a two-photon state characterized by a complex, highly correlated joint-spectral function~\cite{grice}. In our context, the spectral correlations together with the fact that the triggering photon is measured with a conventional frequency-blind APD, lead to an effect akin to dephasing and the heralded photon ends up in a mixed state.  One possibility to produce highly pure single photon states relies on careful engineering of the phase-matching condition of the non-linear material~\cite{uren}. Progress in this direction has been described in several recent experiments~\cite{torres_prl, mosley_prl, christoph_rapid}. Nevertheless, a more common strategy to generate pure states still relies on spectral filtering.

We analyzed the performance of our method considering a single photon state produced by parametric down-conversion. To follow our analysis it is only necessary to know that the joint-spectral function $\phi(\nu_s,\nu_i)$ of the photon pair state is the product of the phase-matching (PM) condition $\Phi(\nu_s,\nu_i)$ and the pump envelope $\alpha(\nu_s+\nu_i)$. The first has a spectral shape determined by a sinc function and a slope $\theta$, the latter is assumed to be a Gaussian. 
\begin{figure}[ht]
\centering
\epsfig{file=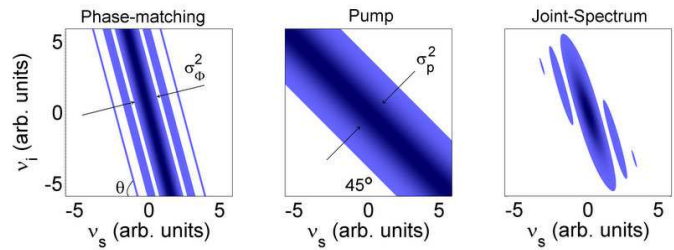,scale=0.5}
\caption{Illustration of the joint-spectral distribution (ellipse), determined from the properties of the pump beam ($\alpha$) and phase-matching ($\Phi$). The standard deviation of the pump spectral function and PM are respectively denoted by $\sigma_p$ and $\sigma_\Phi$. In what follows we will use the notation $w$  to refer to the full width at half maximum (FWHM) of pump (p), phase-matching ($\phi$), spectral filter ($f$) and reference field ($\beta$), which is indicated by the appropriate subscript.}
\label{jointspectrum}
\end{figure}
The parameter $\nu_{s}$ ($\nu_{i}$) denotes the frequency detuning of signal (idler) with respect to the central frequency of the pump. Figure~\ref{jointspectrum} allows to quickly overview the relevant parameters. 
We recall that the spectrum of one field is just the marginal distribution obtained by integrating $\phi$ over the frequencies of the other. Finally, considering that the idler photons are used as a trigger, the spectral density function of signal is given by~\cite{kaisa_optexp}   
\be
g(\nu_s , \nu_s ') \propto \hspace{-3pt} \int d\nu_{i} \ \phi(\nu_s , \nu_{i}) \phi^{*}(\nu_s ', \nu_{i}) \sqrt{f_{s}(\nu_s )}\sqrt{ f_{s}(\nu_s ')}\,f_{i}(\nu_i )  \, ,
\label{gpdc}
\ee
with $f_{\mu}(\nu_{\mu})$ the intensity transmission function of the applied spectral filter.  We note that the joint-spectral function $\phi$ determines the frequency correlations between signal and idler, being a property of two fields. The density function $g$, although connected to $\phi$, is a property of a single beam and has a meaning directly connected to the degree of spectral coherence. Thus care must be taken to avoid confusing these two functions. 

To test our technique, we simulated several conditions, resulting from different choices of the joint-spectral function and eventual use of Gaussian spectral filters. Those define the spectrum and purity of the signal field. Taking a reference beam with similar spectral width we evaluate (\ref{interf}) for several temporal delays, resulting in a specific HOM dip. Fitting Gaussian functions to the fields' spectra and HOM dip, we obtain the necessary parameters to evaluate the purity via (\ref{puritywidth}), which will be denominated as $\mathcal{P}_{\mathrm{width}}$. Finally, we explicitly compare the value obtained from our method with the one found from a direct calculation of the purity as $\mathcal{P}$=Tr$[\rho^2]$ (\ref{purity}).  In what follows we considered the HOM interference of {\it signal} photons, thus $g$ will never refer to the density function of idler.

\subsection{Performance of the method with different wavepackets}
As a general rule, the shape of the HOM dip is dominated by the function with the smallest spectral width. Thus a {\it very} narrow phase-matching (PM) function leads to a squared profile for the temporal dip (Fourier transform of the sinc). If on the other hand the width of the pump field is decreased such that it becomes on the order or narrower than the PM (in frequency units), then the shape of the HOM dip tends to a Gaussian. Below we discuss the main factors to be considered when applying our method, illustrating our proposal with concrete results for three different situations of great interest.

\begin{figure}[ht]
\centering
\epsfig{file=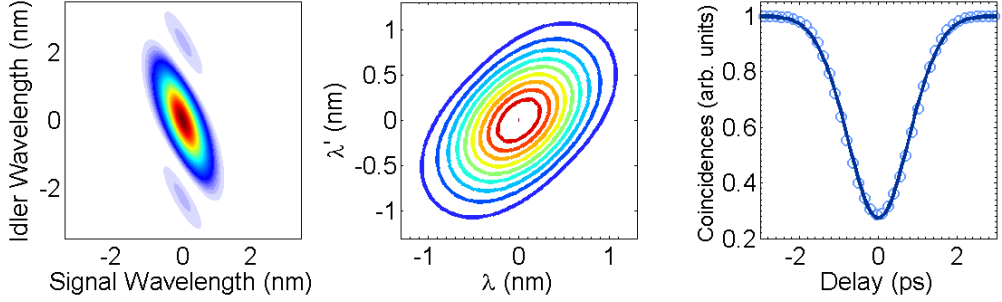,scale=0.55}
\caption{(Left) Filtered highly correlated joint-spectral function. (Middle) Contour lines of $|g(\nu,\nu')|$ as a function of wavelengths. (Right) HOM dip obtained by interfering the signal with the reference field: (circles) calculated using (\ref{interf}), (solid line) Gaussian fit. The following parameters were used: $w_p=2$~nm, $w_{\Phi}=0.8$~nm, $\theta=55^{\circ}$, $w_f=1$~nm and $w_{\beta}=1$~nm.}
\label{gww_pdc}
\end{figure}

(I) Highly frequency correlated joint-spectrum: In this case, to increase the degree of spectral purity of the heralded single photon it is necessary to apply tight filtering. High purity is achieved when the filter bandwidth is on the same order as the intrinsic width of the joint-spectrum (minor axis).  We show in figure~\ref{gww_pdc} (left) a filtered joint-spectral function  that closely resembles the one found in our experiment, which will be discussed later.  
The corresponding spectral density function of the signal $|g(\nu,\nu')|$ is presented in figure~\ref{gww_pdc} (middle). Its shape is very close to a function that is separable along the major and minor axis. Employing the width of the HOM dip, shown in figure~\ref{gww_pdc} (right), one finds the degree of purity estimated with our method: $\mathcal{P}_{\mathrm{width}}=59\%$. This value has to be compared with the spectral purity obtained by the direct calculation: $\mathcal{P}=60\%$. We find good agreement.
\begin{figure}[ht]
\centering
\epsfig{file=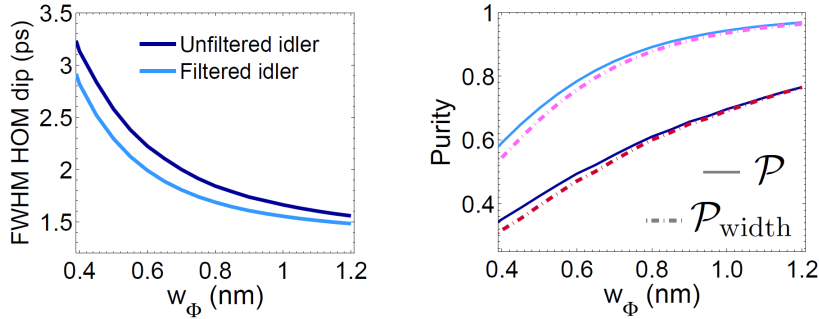,scale=0.45}
\caption{(Left) Width of the HOM dip (defined by the FWHM of a Gaussian fit) for different widths of the phase-matching, considering unfiltered and 1~nm filtered triggering photons. (Right) Comparison between ``true'' purity (solid line) and the one given by our method (dashed line) for unfiltered (dark colours) and 1~nm filtered (light colours) triggering photons. The following parameters were used: $w_p=2$~nm, $\theta=55^{\circ}$, $w_f=1$~nm and $w_{\beta}=1$~nm.}
\label{gww_pdcb}
\end{figure}
In general, to increase  the purity of the signal state even further, a second filter can be included in the idler field. This non-local manipulation of signal's frequency leads to a narrower heralded spectrum, but at the same time to a narrower HOM dip. The latter is a consequence of describing the phase-matching by a sinc function and would not be observed in case it could be approximated by a Gaussian. This effect can be seen in figure~\ref{gww_pdcb} (left), where we present the width of the dip for different PM widths. Thus, by adding a 1~nm filter to limit idler frequencies one finds $\mathcal{P}=89\%$ and $\mathcal{P}_{\mathrm{width}}=88\%$. Again our method proves to be a faithful tool for the characterization of the purity. Its efficacy as a function of the PM width is presented in figure~\ref{gww_pdcb} (right). For small values of the PM width the resulting squared HOM dip profile indicates that the $g_2$ function is not a Gaussian. This leads to retrieving a value for the purity that is about 10$\%$ underestimated. As a general behaviour, the estimate $\mathcal{P}_{\mathrm{width}}$ tends from below towards the correct value of the purity, being optimal when $w_{\Phi} \approx w_f$.

\begin{figure}[ht]
\centering
\epsfig{file=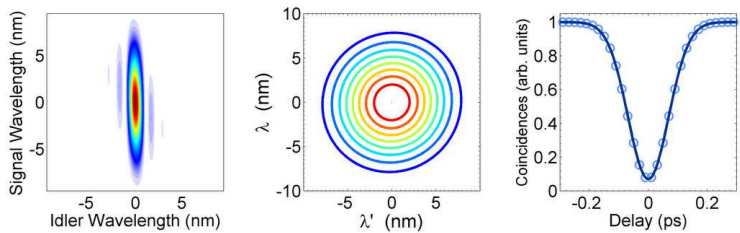,scale=0.55}
\caption{(Left) Uncorrelated joint-spectral function. (Middle) Contour lines of $|g(\nu,\nu')|$ as a function of wavelengths.  (Right) HOM dip obtained by interfering signal with the reference field: (circles) calculated using (\ref{interf}), (solid line) Gaussian fit. The following parameters were used: $w_p=1.8$~nm, $w_{\Phi}=1$~nm, $\theta=90^{\circ}$ and $w_{\beta}=10$~nm.}
\label{gww_pdc2}
\end{figure}

(II) Phase-matching lying at 90$^{\circ}$: High purity heralded single photons are generated by appropriate choice of the pump width and there is no need for spectral filtering. A typical joint-spectrum and the corresponding density function of the heralded photon are illustrated  in figure~\ref{gww_pdc2}(left) and (middle), respectively. In this case, by tracing over idler frequencies to obtain $\rho_s$ one naturally introduces a smearing out of the sinc structure, resulting in a  $g(\nu,\nu')$ function that is a 2-dimensional Gaussian. Therefore, our method can be used without restrictions and we find that $\mathcal{P}=\mathcal{P}_{\mathrm{width}}=96\%$.

\begin{figure}[ht]
\centering
\epsfig{file=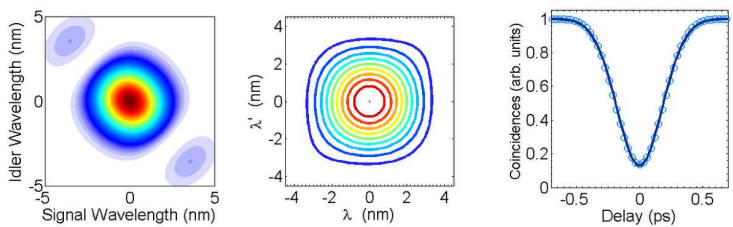,scale=0.55}
\caption{(Left) Uncorrelated joint-spectral function. (Middle) Contour lines of $|g(\nu,\nu')|$ as a function of wavelengths. (Right) HOM dip obtained by interfering signal with the reference field: (circles) calculated using (\ref{interf}), (solid line) Gaussian fit. The following parameters were used: $w_p=1$~nm, $w_{\Phi}=3$~nm, $\theta=-45^{\circ}$ and $w_{\beta}=4$~nm.}
\label{gww_pdc3}
\end{figure}

(III) Phase-matching lying at 45$^{\circ}$: Again high purity is achieved by matching the widths of pump and the PM, resulting in a condition that is the most unfavorable for our technique. The reason is that the ``wings'' of the sinc function appear in a pronounced and symmetric way, thus surviving to the smearing effect discussed above. We show in figure~\ref{gww_pdc3}(left) and (middle) a typical joint-spectrum and density function, respectively. 
Although the function $g(\nu,\nu')$ is not separable and the validity of the method breaks, a blind application of (\ref{puritywidth}) results in a characterization of the purity that is only about 12$\%$ wrong, nevertheless in this case it is overestimated. This state has a purity of $\mathcal{P}=88\%$ and, given its reduced value, the use of traditional spectral filters is still inevitable. The local application of a Gaussian spectral filter to the single photon state effectively smooths the effect of the sinc bringing the spectral density function towards the ideal conditions of our method. Applying a 5~nm filter to the signal field, enough for cutting most of the ``wings'' arising from the sinc function, would result in $\mathcal{P}=96\%$ and  a 4$\%$ overestimated $\mathcal{P}_{\mathrm{width}}$.

\section{Experiment}
\label{secexperiment}
\begin{figure}[ht]
\centering
\epsfig{file=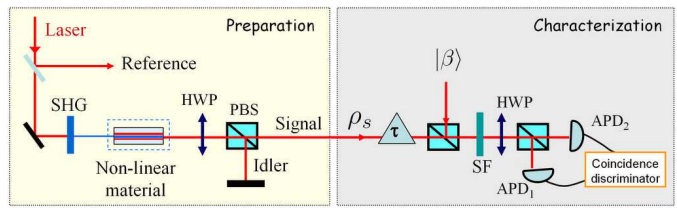,scale=0.5}
\caption{Sketch of the experimental setup. SHG: second harmonic generation, SF: spectral filter ($w_f=1.0(1)$~nm), HWP: half-wave plate,  PBS: polarizing beam-splitter, APD: Avalanche photodetector. }
\label{setup}
\end{figure}

To show the experimental feasibility of the method, we measured the HOM dip between one of the twins and a coherent reference. An illustration of the experimental setup is presented in figure~\ref{setup}. A 1.45~mm long, type-II, periodically poled KTiOPO$_4$ waveguide is pumped by a frequency doubled Ti:Sapphire laser operating in the ultrafast regime (796~nm, FWHM equals to 10~nm, autocorrelation length of 170~fs and repetition rate of 4~MHz). A small fraction of the fundamental light is used as the reference field. The spectral width of the pump field is $w_p=2.0(1)$~nm.
The path delay $c \, \tau$ between reference and one of the twins is adjusted by using a high-precision translation stage, which is moved in steps corresponding to a delay of $0.22(1)~$ps. The fields are combined by using a fibre integrated beam-splitter, thus ensuring a good spatial overlap. We measure coincident clicks using two avalanche photodetectors whose signals are acquired by a computer using a time-to-digital converter (TDC).

In our case the joint-spectrum is highly correlated and we prepare heralded single photons by applying 1.0(1)~nm spectral filters to both, signal and idler photons. Discarding any information about idler ({\it unconditional} measurement), it is legitimate to assume that the spectral width of the signal is equal to the filter width  $\sigma_{g1}=\sigma_f$, since the unfiltered marginal spectrum is much broader. A similar condition holds for the reference beam, $\sigma_{\beta}=\sigma_f$. Thus, to avoid measuring  $\sigma_{g1}$, we begin the experiment by investigating the interference between the unconditional signal state and the reference, i.e. between a thermal and coherent state.

\begin{figure}[ht]
\centering
\epsfig{file=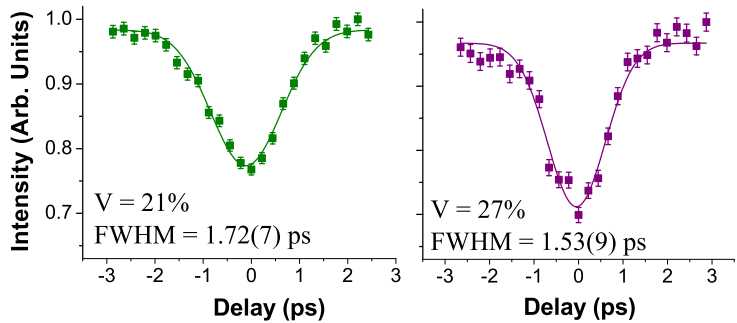,scale=0.35}
\caption{Measured HOM dip by interfering the reference field with: (Left) signal, (Right) idler. The width of the dip and the corresponding visibility are shown in the plots.}
\label{results}
\end{figure}

The result of our measurement is presented in figure~\ref{results}, where we show  typical HOM dips observed by interfering either signal or idler with the reference beam. By analyzing several similar sets of data we extract the following FWHM values: $w_\mathrm{s}=1.75(5)$~ps and $w_\mathrm{i}=1.54(5)$~ps. The purity of the unconditional mode is retrieved applying (\ref{puritywidth}). One finds that $\mathcal{P}^{(s)}_{\mathrm{width}}=0.64(10)$ for the signal and, in a similar way, $\mathcal{P}^{(i)}_{\mathrm{width}}=0.77(10)$ for the idler, despite the low visibilities (depicted in figure~\ref{results}).

\begin{table}[ht]
\begin{center}
\begin{tabular}{|ccccc|}
\hline\hline
 & \;$p_0$ ($\%$) \;& $p_1$ ($\%$)  \; & $p_2$ ($\%$) &  $|\beta|^2$  \\
\hline
Signal & 99.7015   & 0.2978  &  0.0006  &  0.0030 \\
Idler & 99.8309   & 0.1689  &  0.0002  &  0.0027\\
\hline
\end{tabular}
\caption{ Mean photon number of the reference field and photon number statistics of the field that interferes with it.}
\label{pnd}
\end{center}
\end{table}
The best way to compare our method with the standard one based on the visibility is to analyze the value of the overlap factor, since for the interference between two intrinsically different sources the visibility is not {\it directly} connected to the purity~\cite{mosley_njp}.  As discussed, this quantity can be retrieved either from the HOM width via (\ref{specoverlap}) or from the visibility via (\ref{visibility}). Here we refer to the former  as $T$ and to the latter as $T_{\mathrm{vis}}$. For the first case we directly obtain $T^{(s)}=0.76$ and $T^{(i)}=0.86$. For second case, we measured the photon number distribution of both fields using a standard time-multiplexing detector~\cite{kaisa_optexp}. In our experiment we can neglect the effect of dark counts by applying a tight temporal gating to the acquisition ($\sim$3~ns). The observed probabilities of vacuum $p_0$, one-photon $p_1$ and two-photon $p_2$ components of the studied state, together with the mean photon number $|\beta|^2$ of the coherent field, are shown in Table~\ref{pnd}. With these values one can calculate $\mathcal{S}=p_1\,|\beta|^2$ and $\mathcal{B}=[p_0 \,|\beta|^4/2 + p_2]$, which yield an overlap factor of $T^{(s)}_{\mathrm{vis}}=0.46$ and $T^{(i)}_{\mathrm{vis}}=0.60$, smaller than the ones retrieved from the width of the dip.  This exemplifies well the fact that the visibility is affected by mode mismatch and other experimental imperfections, contrarily to the temporal width used in our method.

Concerning the heralded photon state, the non-local spectral filter in the idler arm  reduces the spectral width of the signal state $\sigma_{g1}$, thus differing from $\sigma_{f}$ (bandwidth of the filter in the signal arm). Therefore, in this scenario $\sigma_{g1}$ has to be measured in order to allow the characterization of the heralded purity via the HOM width. Unfortunately, in our case this width is narrower than 1~nm and can not be directly accessed in our system. Still, the degree of purity can be estimated making use of the knowledge that our source produces photon pairs, described by a joint-spectral function, with measured unconditional purities. In previous measurements we characterized the slope of the phase-matching function to be $\theta=54.7^{\circ}$~\cite{kaisa_optexp}, thus by comparing the values of $\mathcal{P}^{(s)}_{\mathrm{width}}$ and $\mathcal{P}^{(i)}_{\mathrm{width}}$ with a theoretical curve (obtained by numerical integration), we can characterize the PM width. From this analysis we estimate that, employing a 1~nm filter in the idler arm, the heralded purity is $\mathcal{P}_h=0.90(6)$ and the spectral overlap  with respect to the same 1~nm broad reference field is $T=0.88$.

\section{Conclusion}
We demonstrated a new method to directly determine the purity of single photons in a way that is independent from mode matching. 
We employ the standard Hong-Ou-Mandel interference between two fields but, contrarily to the usual approach based on the amplitude of the interference, we benefit from the robustness of its temporal width against mode mismatch. Its applicability for characterizing the state produced by photon pair sources is explicitly examined and discussed for different experimental scenarios. In particular, we show its advantage in an experiment dealing with tightly filtered parametric down converted photons. 

Our technique can also be extended to situations in which signal and reference with the same wavelength are not easily  available. In this case, the solution might be to employ two independent but equivalent photon pair sources and interfere the signal mode from each of them. The information about the purity should still be available in the temporal width of the HOM dip.

We hope that our work will trigger further investigations about more practical and robust techniques for quantifying the quality of general single photon states.
We believe that our ideas will boost the characterization and optimization of high-purity heralded single photon states, thus having direct impact on the practical realization of quantum information protocols~\cite{obrian, obrian2, briegel}.

\section*{Appendix}
The precise expressions used in our simulations are provided below. The phase-matching is defined as
\be
\Phi(\omega_{s},\omega_{i}) = \textrm{sinc}(\Delta k(\omega_{s},\omega_{i})L/2)e^{i \Delta k(\omega_{s},\omega_{i})L/2} \,,
\nonumber
\ee
where $\{\omega_{s},\omega_{i}\}$ represents the frequency space of signal and idler fields, $\Delta k$ the phase-mismatch between pump and twins and $L$ the length of the nonlinear medium. We define $\nu_{\mu}$ ($\mu=s,i,p$) as the detuning  relative to the central frequencies of signal, idler, and pump, respectively. Moreover, the first order expansion of the phase-mismatch reveals that $\Delta k(\nu_{s},\nu_{i})=\kappa_{s}\nu_{s} +  \kappa_{i}\nu_{i}$, where $\kappa_{\mu}$  is the respective group velocity mismatch with respect to the pump. With this consideration the PM function is rewritten as
\be
\Phi(\nu_{s},\nu_{i}) \approx  \textrm{sinc}\left( \frac{\sin\theta \nu_{s} +  \cos\theta \nu_{i}}{\sqrt{2\gamma \sigma_\Phi^{2}}}\right) e^{i\frac{L}{2}(\kappa_{s}\nu_{s} +  \kappa_{i}\nu_{i})} \,,
\nonumber
\label{phasematching}
\ee
with $\tan\theta=\kappa_{s}/\kappa_{i}$, $\sigma_\Phi^{-2}=(\kappa_{s}^2+\kappa_{i}^2)\gamma L^2/2$, and $\gamma=0.193$. Since directly measurable observables are proportional to the intensity, we define the phase-matching width $w_{\Phi}$ by the FWHM of $|\Phi|^2$, i.e. $w_{\Phi}=2\,\sigma_\Phi\,\sqrt{\ln{2}}$. 

The pump field is considered to have a Gaussian spectral profile, defined by
\be
\alpha(\nu_{s},\nu_{i}) = \exp(-(\nu_{s} + \nu_{i} )^{2}/(2 \sigma_p^{2}))\, ,
\nonumber
\ee
with FWHM (intensity) given by  $w_{p}=2\,\sigma_p\,\sqrt{\ln{2}}$. Finally, using (\ref{phasematching}) and (\ref{gpdc}) we note that the linear phase term of $g(\nu_s , \nu_s ')$ is $[L/2\;\kappa_s\,(\nu_s-\nu_s')]$. 

In part of our analysis we employed spectral filters. Those are defined by Gaussian functions $f(\om)$, such that $f(\om)^2$ has standard deviation and FWHM respectively denoted by $\sigma_f$ and  $w_f$.  Moreover, for the characterization of the purity we assumed a reference field with Gaussian spectral profile $u(\om)$. The standard deviation of $u(\om)^2$ is denoted by $\sigma_{\beta}$ and the respective FWHM by $w_{\beta}$.

\ack
This work was supported by the EC under the grant agreement CORNER (FP7-ICT-213681).
K.N.C. acknowledges financial support from the Alexander von Humboldt Foundation, and helpful discussions with A. S. Villar, C. S\"oller and P. Nussenzveig.


\end{document}